# Minimizing the Value-at-Risk of Loan Portfolio via Deep Neural Networks

**Di Wang** [a,b] and **Ye Du** [a, b]

[a] Department of Financial Engineering, School of Finance
[b] Collaborative Innovation Center for the Innovation and Regulation of Internet-based-Finance
Southwestern University of Finance and Economics
Chengdu, Sichuan, China
{ albertwang0921, henry.duye }@gmail.com

## Abstract

Risk management is a prominent issue in peer-to-peer lending. An investor may naturally reduce his risk exposure by diversifying instead of putting all his money on one loan. In that case, an investor may want to minimize the Value-at-Risk (VaR) or Conditional Value-at-Risk (CVaR) of his loan portfolio. We propose a low degree of freedom deep neural network model, DeNN, as well as a high degree of freedom model, DSNN, to tackle the problem. In particular, our models predict not only the default probability of a loan but also the time when it will default. The experiments demonstrate that both models can significantly reduce the portfolio VaRs at different confidence levels, compared to benchmarks. More interestingly, the low degree of freedom model, DeNN, outperforms DSNN in most scenarios.

## 1 Introduction

Peer-to-Peer (P2P) lending is a booming market in the financial industry. For instance, Lending Club, one of the largest peer-to-peer online lending platforms in the world, has funded more than ten billion dollars in loans during the past few years. Compared with lending/borrowing through the traditional banking system, this business model significantly reduces the transaction cost incurred by financial intermediaries and therefore rewards lenders [1] with higher returns and borrowers with lower costs. In P2P lending, a borrower can issue a loan to finance from multiple lenders and make the promise to pay the monthly installments during the *loan term* that is the lifetime of a loan if it would not default.

However, standing in the lenders' shoes, one natural concern is the possible defaults of loans. Risk management becomes a prominent issue in peer-to-peer lending. In order to reduce his own risk, a lender may scrutinize each loan carefully and put his money on the safest asset he believes. However, given the limited resources of an average lender and the large amount of loans in a P2P lending platform, it is not efficient at all for himself to do so. Thus, he may diversify his investment by lending to multiple borrowers, i.e., holding a portfolio of loans. How to minimize the risk of an investor in this scenario is the focus of our study.

The risk of investing a portfolio of loans is measured by Value-at-Risk (VaR), which is a standard risk measure in finance and advocated by Basel II Accord [Jorion, 2007]. Essentially, the VaR of an asset is the negative value of a certain percentile of the asset's return distribution. Therefore, to compute the VaR of a loan portfolio, return distributions of its constituent loans must be known. In this paper, two methods based on deep neural networks are proposed to predict the distribution of return for each loan. In the Default Neural Networks (DeNN) method, the return of a loan is modeled as a binary random variable, i.e., the realized return under default and the realized return under non-default. In the Deep Survival Neural Network (DSNN) method, the return of a loan is modeled as a random variable with multiple individual realizations for the default at each installment period. In terms of model structure, the DeNN method employs a neural network (DL-NN) to directly predict the default lifetime of a loan, while it uses another neural network (DR-NN) to predict the default rate. The distribution of return is obtained by combining the outputs of these two neural networks. By contrast, the DSNN method models the default lifetime with a higher degree of freedom in the sense that it has more variables to predict. The DSNN is a novel two-branch deep neural network with specially designed loss and structure. After training, one branch of DSNN is extracted to predict the default rate of a loan and the distribution of its default lifetime, which induces the distribution of return. Following the prediction of return distribution, we use Monte-Carlo methods to simulate possible realized returns for loans in a portfolio, assuming loans' returns are independent of each other. Finally, the optimal weights invested in loans are generated to minimize the simulated VaR of the portfolio.

The contributions of this paper are as follows. Although there are many studies about default rate prediction of a loan, to the best of our knowledge, our work is the first to use deep neural networks to predict the return distributions of P2P loans. It further makes this paper the first one in literature to tackle the problem of loan portfolio VaR optimization. Indeed, our two models, DeNN and DSNN, can significantly reduce the VaRs at different confidence levels, which is demonstrated through experiments on real-life P2P loan data from Lending Club. Last but not the least, the DeNN model, which has a low degree of freedom and is quite simple in design, outperforms the DSNN model in most scenarios, which has a relatively high degree of freedom and is sophisticated in design. This result is interesting and could be insightful in guiding the use of neural networks in practice.

---

[1] In this paper, lender and investor are used interchangeably.

## 2 Two Methods for Loan Return Distribution

In this section, we introduce the DeNN method and DSNN method for the prediction of loan return distribution. For better illustration, Table 1 lists the mathematical notations and explanations for some terminologies used in this paper.

| Notation | Description |
|---|---|
| E | Default Indicator: a loan will default if insufficient times of repayments are made; E=1 if a loan defaults, otherwise 0. |
| M | Loan Amount: it is the amount of money the borrower financed. |
| C | Installment: loans are repaid monthly; installment is the amount of money needed to repay per month; it is same for every month. |
| $\tilde{t}$ | Lifetime: it is the number of months between loan issuance and its last repayment. |
| L | Loan Term: it is the lifetime of a loan if it would not default; it equals the arranged number of repayments. |
| $\tilde{t}_d$ | Default Lifetime: it is the lifetime of a loan if it would default; it equals the number of repayments made. |
| $r_L$ or $\tilde{r}_d$ | Loan Return: it is the return rate of a loan; it equals the promised return rate $r_L$ if the loan would not default; if a loan would default, its return rate ($\tilde{r}_d$) can be computed using Eq. (1). |

Table 1: Mathematical Notations

In finance, Eq. (1) is the standard formula to compute the loan return, (note that $\tilde{r}_d$=-1 when $\tilde{t}_d$=0).

$$M = \sum_{t=1}^{\tilde{t}_d} \frac{C}{(1 + \tilde{r}_d)^t} \qquad (1)$$

### 2.1 DeNN: Default Neural Networks

In this part, the DeNN method is introduced to predict the return rate distribution of a loan. According to the definition of loan return, the distribution of loan return can be obtained once the default rate (denoted by $p$) and the default lifetime $\tilde{t}_d$ are predicted. Here, two deep neural networks are employed to address these two tasks separately.

**Neural Network for Default Rate**
To predict the default rates of loans, a feedforward fully-connected five-layer neural network (named DR-NN) is built. The DR-NN has 104 neurons in the input layer and 1 neuron in the output layer. The number of neurons in three hidden layers are 128, 64, and 32, respectively. The activation functions are 'tanh' in the hidden layers and 'sigmoid' in the output layer.

The inputs of DR-NN are the features of a loan, and the output is the predicted default rate $\hat{p}$. The output is supervised by the default indicator E during training. The loss is the binary cross entropy,

$$loss_{DR-NN}(\boldsymbol{\theta}) = -\frac{1}{b}\sum_{i=1}^{b}(E_i \cdot \ln \hat{p}_i + (1 - E_i) \cdot \ln(1 - \hat{p}_i)), \qquad (2)$$

where $i$ denotes the loans in a batch, $b$ is the batch size in the mini-batch gradient descent algorithm, and $\boldsymbol{\theta}$ is the parameters of DR-NN. Besides, during training, an L2 regularization term is added to Eq. (2) to improve the generalization ability. A weighted loss is employed to deal with the data imbalance problem [Bishop, 2007; Goodfellow et al., 2016].

**Neural Network for Default Lifetime**
Another neural network (named DL-NN) that has the same structure and activation functions as DR-NN is built to predict the default lifetimes of loans. The inputs of DL-NN are also the features of a loan, and the output is the predicted default lifetime ratio that is the value of default lifetime divided by loan term. It is supervised by the observed default lifetime ratio. The loss is the mean squared error (MSE). During training, just default loans are used because only they have the label, i.e., the observed default lifetime ratio. Also, an L2 regularization term is added to the MSE loss to improve the generalization. After training, DL-NN can be used to predict the default lifetime by multiplying its output by loan term L.

In summary, we model the loan return $\tilde{r}$ as a binary random variable as follows,

$$\tilde{r} = \begin{cases} r_L, & 1 - \hat{p} \\ \hat{r}_d, & \hat{p} \end{cases}, \qquad (3)$$

where $r_L$ is known, $\hat{p}$ is the predicted default rate, which is the output of DR-NN, and $\hat{r}_d$ is the predicted value of $\tilde{r}_d$. $\hat{r}_d$ is computed by plugging $\hat{t}_d$, the prediction of DL-NN, into Eq. (1).

### 2.2 DSNN: Deep Survival Neural Network

In this part, a novel two-branch deep neural network with specially designed loss and structure (named DSNN) is designed to predict the distribution of default lifetime. To illustrate, a loan default may take place at each installment period. While the DeNN method focuses on estimating the default lifetime as accurate as possible for a loan, the DSNN focuses on outputting the distribution of the default lifetime, i.e., the probabilities of default at $\tilde{t}_d$=0, 1, ..., or L-1,

$$\forall\, 0 \le i \le L-1,\ Pr(\tilde{t}_d = i) = p_i. \qquad (4)$$

According to the definition of loan return, if a loan would default, its return is solely determined by its default lifetime.

By contrast, if it would not default, the return is $r_L$ and the probability of non-default can be computed by Eq. (5).

$$1 - p = 1 - \sum_{i=0}^{L-1} p_i \qquad (5)$$

Therefore, the distribution of loan return can be obtained once the distribution of default lifetime is predicted.

**Survival Neural Network Branch**
Survival Analysis is a standard framework in statistics to model the duration of time until a certain event happens [Kleinbaum and Klein, 2010]. Here, in order to predict the distribution of the default lifetime for a loan, we build a feedforward fully-connected five-layer neural network (named SNN) to first estimate its hazard function and survival function. This follows a similar design of network architecture and loss function, which has been used for equipment health management [Liao and Ahn, 2016].

The hazard function has the following form,

$$h(t) = \lambda^\rho \rho t^{\rho-1} exp(SNN_\theta(\pmb{x})), \qquad (6)$$

where $t$ is the lifetime, $\lambda$ and $\rho$ are the parameters of the Weibull distribution fitted by all loan's lifetime, $SNN_\theta(\pmb{x})$ represents the output of the SNN, and $\theta$ is the parameters of SNN. The corresponding survival function is

$$S(t) = exp\left(-(\lambda t)^\rho \cdot e^{SNN_\theta(\pmb{x})}\right). \qquad (7)$$

The SNN has the same structure as DR-NN. Its activation functions are 'tanh' in the first two hidden layers, 'sigmoid' in the last hidden layer, and 'linear' in the output layer. The loss function for SNN is the negative log likelihood as follows,

$$loss_{SNN}(\pmb{\theta}) \coloneqq -\frac{1}{b} \sum_{i=1}^b \{E_i \cdot \ln[h(t_i|\pmb{x}_i)] + \ln[S(t_i|\pmb{x}_i)]\}, \qquad (8)$$

where $i$ and $b$ remain the same meaning as in Eq. (2), and $t_i$ is the observed lifetime. After training SNN, the predicted survival function is formulated by Eq. (7). According to the definition of survival function, the probabilities in Eq. (4) can be computed as

$$p_i = \begin{cases} 1 - S(0), & i = 0 \\ S(i-1) - S(i), & 1 \le i \le L-1 \end{cases} \qquad (9)$$

Besides, the non-default rate is $S(L-1)$. In this way, the default lifetime distribution and default rate are predicted.

**Deep Neural Network Branch**
In literature, various machine learning techniques are applied to predict the default rates of loans [Yang, 2007; Zhang *et al.*, 2014; Byanjankar *et al.*, 2015; Malekipirbazari and Aksakalli, 2015; Zhao *et al.*, 2016; Xia *et al.*, 2017]. These techniques achieve great performance on this binary classification task. While the task of default rate prediction is relatively simple as it only has two dimensions in the label space, the task of the default lifetime distribution prediction is much more complicated.

However, considered that the default rate of a loan can also be derived from its predicted survival function (as Eq. (5) and Eq. (9) suggests), one natural idea is to supervise the default rate derived from SNN's prediction by the default rate prediction from an 'expert', i.e., a deep neural network trained exclusively to predict the defaults of loans. By doing so, the SNN is supposed to achieve better results in the prediction of default lifetime distribution. Here, the 'expert' (named DNN) has the same structure and loss as the DR-NN.

**The Connected Neural Network**
According to Eq. (5) and (9), the predicted default rate from SNN is

$$\hat{p}_{SNN} = 1 - S(L-1). \qquad (10)$$

Plugging in the specific form of survival function in Eq. (7), we get

$$\hat{p}_{SNN} = 1 - exp\left(-\left(\lambda \cdot (L-1)\right)^\rho \cdot e^{SNN_\theta(x)}\right). \qquad (11)$$

Therefore, a neuron (named N1) that does the computation in Eq. (11) is designed to receive the output of SNN. Then, since $\hat{p}_{SNN}$ should coincide with the predicted default rate of DNN, $\hat{p}_{DNN}$, the outputs of N1 and DNN are further passed to another neuron, N2, which computes the distance between $\hat{p}_{SNN}$ and $\hat{p}_{DNN}$. The distance is represented by the square of the difference between them. Finally, this distance is supervised by value zero. This process can be summarized as the following loss function.

$$loss_{dif} \coloneqq \frac{1}{b} \sum_{i=1}^b \left(\hat{p}_{DNN,i} - \hat{p}_{SNN,i}\right)^2, \qquad (12)$$

where $i$ and $b$ remain the same meaning as in Eq. (2). In this way, we connect SNN and DNN to build the DSNN. The loss used to train DSNN is

$$loss(\pmb{\theta}) = w_{SNN} loss_{SNN} + w_{DNN} loss_{DNN} + w_{dif} loss_{dif}, \qquad (13)$$

where $\theta$ is the parameters of DSNN. $w_{SNN}$, $w_{DNN}$, and $w_{dif}$ are the parameters to distinguish the importance of different losses. In this study, these three parameters are set as 1. After training, the SNN branch is extracted to predict the survival function.

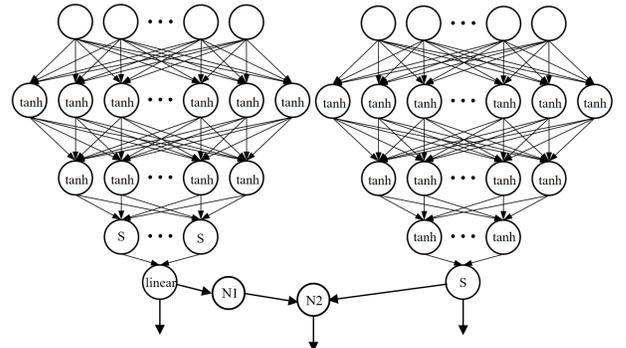

Figure 1: Deep Survival Neural Network

Figure 1 shows an illustrative structure of DSNN. Three outputs are supervised by $loss_{SNN}$, $loss_{DNN}$, and $loss_{dif}$, respectively. The distribution of loan return is denoted as,

$$\tilde{r} = \begin{cases} r_L, & S(L-1) \\ r_{d,i}, & p_i, \forall\ 0 \le i \le L-1 \end{cases} \quad (14)$$

where $p_i$ can be computed by Eq. (9), and $r_{d,i}$ is the return when $\tilde{t}_d = i$, which can be computed by Eq. (1).

## 3 VaR Minimization via Simulations

Hitherto, the distribution of return for each loan is predicted by DeNN or DSNN method. In this section, given that $n$ loans are available to invest in, we propose a Monte-Carlo simulation based algorithm to minimize the VaR and CVaR of this $n$-loan portfolio.

### 3.1 Monte-Carlo Simulations

For $n$ loans available to invest in, a portfolio is constructed by allocating your capital among them. Given that the allocation weights are $\boldsymbol{w} = [w_1, w_2, ..., w_n]$, where $0 \le w_i \le 1$, the return rate of this $n$-loan portfolio can be expressed as follows,

$$\tilde{r}_p = \sum_{i=1}^{n} \tilde{r}_i, \quad (15)$$

where $i$ denotes the loans in the portfolio, $n$ is the size of portfolio, $\tilde{r}_p$ is the portfolio return, and $\tilde{r}_i$ is the loan $i$'s return.

While the possible values of $\tilde{r}_i$ can be presented in the form of a set, such as $H_i = \{r_d, r_L\}$ for DeNN method or $H_i = \{r_{d,0}, r_{d,1}, ..., r_{d,L-1}, r_L\}$ for DSNN method, the set of possible return rates of $n$ loans in a portfolio can be expressed by a $n$-array Cartesian product,

$$H_p = H_1 \times H_2 \times \cdots \times H_n, \quad (16)$$

where the set $H_p$ has $m$ elements ($m$=$2^n$ for DeNN method, $m$=$(L+1)^n$ for DSNN method), representing $m$ scenarios of a loan portfolio. Then, all realized values of $\tilde{r}_p$ can be represented by a vector, $R_p$.

$$R_p = \begin{bmatrix} r_{p,1} \\ r_{p,2} \\ \vdots \\ r_{p,m} \end{bmatrix}^{\mathrm{T}} = \begin{bmatrix} w_1 \\ w_2 \\ \vdots \\ w_n \end{bmatrix}^{\mathrm{T}} \begin{bmatrix} r_{1,1} & r_{1,2} & \cdots & r_{1,m} \\ r_{2,1} & r_{2,2} & \cdots & r_{2,m} \\ \vdots & \vdots & \ddots & \vdots \\ r_{n,1} & r_{n,2} & \cdots & r_{n,m} \end{bmatrix} \quad (17)$$

In Eq. (17), $r_{p,j}$ is the portfolio return in scenario $j$, and $r_{i,j}$ is the return of loan $i$ in scenario $j$. The last term in Eq. (17) is a return matrix denoted by $R_{n \times m}$. Because the exhaustive enumeration of $m$ scenarios is impossible, a simulated return matrix $R_{n \times k}$ is generated by Monte-Carlo simulations to replace $R_{n \times m}$.

$$R_{n \times k} = \begin{bmatrix} r_{1,1} & r_{1,2} & \cdots & r_{1,k} \\ r_{2,1} & r_{2,2} & \cdots & r_{2,k} \\ \vdots & \vdots & \ddots & \vdots \\ r_{n,1} & r_{n,2} & \cdots & r_{n,k} \end{bmatrix} \quad (18)$$

Specifically, $n$ loans' return rates (a column in $R_{n \times k}$) are simulated at each time and we simulate $k$ times. At each time of simulation, every element of a column in $R_{n \times k}$ is independently drawn out from its corresponding predicted loan return distribution. Correlations between any two loan returns are ignored [Zhao *et al.*, 2014; Guo *et al.*, 2016]. The portfolio return is uniformly distributed on the simulated returns, $\boldsymbol{w}^{\mathrm{T}} R_{n \times k}$. In this study, $k$ is set as 10,000.

### 3.2 VaR Minimization

The risk of a portfolio is measured by Value-at-Risk (VaR) with two parameters: confidence level $\alpha$ and time horizon $l$. For example, a 10-day VaR at 95% confidence level ($\alpha$=95%, $l$=10) equals 27% means that with only a probability of 5%, we can expect a return lower than -27% if we buy and hold the portfolio over the next 10 days.

Essentially, the $l$-day portfolio VaR at $\alpha$ confidence level is just the negative value of the $100 \times (1-\alpha)$ percentile of the portfolio's $l$-day return distribution. The VaR is a highly compressed measure of risk that answers the frequently asked question: under $\alpha$ circumstances, what is the most I will lose on this investment in the next period. It highlights the potential loss of our investment under bad situations. In this paper, time horizon $l$=30 days because monthly return rate is used as our profitability measure. Based on portfolio return distribution simulated in Section 3.1, the VaR at $\alpha$ confidence level is defined as

$$\text{VaR}_\alpha(\boldsymbol{w}) = -Percentile_{1-\alpha}(\boldsymbol{w}^{\mathrm{T}} R_{n \times k}). \quad (19)$$

The $Percentile_{1-\alpha}(\cdot)$ function in Eq. (19) first sorts the elements in vector $\boldsymbol{w}^{\mathrm{T}} R_{n \times k}$ from low to high and then output the value at position $\lfloor (1-\alpha) \times k \rfloor$.

In fact, although $\text{VaR}_\alpha$ measures the maximum loss under $\alpha$ circumstances, it ignores the loss in the rest $(1-\alpha)$ scenarios. To control the potential loss in this situation, another risk measure, CVaR, is used in this study. The $\text{CVaR}_\alpha$ is the expected value of $\text{VaR}_x$ for all $x$ larger than $\alpha$.

$$\text{CVaR}_\alpha(\boldsymbol{w}) = \text{E}(\text{VaR}_x(\boldsymbol{w})|x \ge \alpha) \quad (20)$$

As Eq. (15) suggests, an individual investor can reform the return distribution of his portfolio by choosing different weights, which further results in different VaR/CVaR values. Thus, instead of simply diversifying his portfolio on as many loans as possible, he can directly minimize the portfolio VaR/CVaR to manage the risk of his investment, especially the risk at the tail of portfolio return distribution where bad situations happen. This can be formulated as an optimization problem.

$$\min_{\boldsymbol{w}=[w_1,w_2,...,w_n]} \text{VaR}_\alpha(\boldsymbol{w}) \text{ or } \text{CVaR}_\alpha(\boldsymbol{w})$$
$$\text{subject to } \begin{cases} w_i \in [0,1] \\ \sum w_i = 1 \end{cases} \quad (21)$$

The problem in Eq. (21) is a multivariable optimization problem with a non-convex objective function. The Sequential Least Squares Programming (SLSQP) algorithm [Kraft, 1988] is used to solve it.

## 4 Experiments

In this section, extensive experiments are designed to demonstrate the effectiveness of our methdos. Experiments are implemented using Python. The Pandas and Numpy libraries are used to process data. The Keras library are used to build, train, and test neural networks. And the Scipy library is used to implement the optimization algorithm.

### 4.1 Description of Dataset

The dataset was downloaded from Lending Club, one of the largest P2P online lending platforms in the world. After thoroughly cleansing and preprocessing, 244,720 loans with 104 features and 2 labels (lifetime and default indicator) are left. These features collaboratively describe various aspects of a loan, including a loan's basic information (loan amount, grade, installment, etc.) and the borrower's basic information (income, credit score, home ownership, etc.). The loan term of the 244,720 loans is 36 months. The loans are randomly split into three parts: training set (164,720 loans), validation set (40,000 loans), and test set (40,000 loans).

### 4.2 Results of VaR Minimization

Given $n$ loans in a portfolio, once their return distributions are predicted, the optimal weights can be generated by VaR minimization through Monte-Carlo simulations. Then, the realized portfolio return can be acquired by aggregating the realized returns of its constituent loans (as Eq. (15) suggests). However, only one realized portfolio return can be observed, which means the VaR of this portfolio is not directly accessible. Here, we come up with a solution to deal with this problem. We set the portfolio size $n$=40, and randomly sample 2,000 portfolios with replacement from the test set. Since the 40 loans in each portfolio are randomly selected, we could assume that these 2,000 portfolios are homogeneous and regard them as 2,000 different scenarios for a virtual representative portfolio P. From this point of view, once the capital allocation weights for each portfolio are given, the realized returns of the 2,000 portfolios can be computed using 40 realized loan returns in each portfolio. Then, the 2,000 realized returns can be regarded as the realized returns of the virtual representative portfolio P under 2,000 different scenarios. In this way, the VaR of the portfolio P can be computed straightforwardly.

In experiments, four methods for loan return distribution prediction, and two baselines are compared.

- **DeNN**: the DeNN method proposed in Section 2.1;
- **LRM**: the benchmark for DeNN; using logistics regression for default rate prediction and rating-based model [Guo *et al.*, 2016] for default lifetime prediction;
- **DSNN**: the DSNN method proposed in Section 2.2;
- **SNN**: the benchmark for DSNN; training only the SNN to predict the survival function [Liao and Ahn, 2016].
- **Equal**: equal weights; allocate capital equally;
- **Random**: random weights; allocate capital randomly.

As little previous literature focuses on minimizing the VaR/CVaR of a loan portfolio, there is no baseline model from previous works. Given that simple diversification can be a very efficient way to control risk, random and equal weights are used as two baselines for this task. Since the VaR at 95% and 99% confidence levels are most widely used in finance, we produce the results for four objectives in optimization, i.e., optimizing $VaR_{95\%}$, $VaR_{99\%}$, $CVaR_{95\%}$, and $CVaR_{99\%}$. The results are shown in Table 2-5.

The VaRs in Table 2-5 are annualized losses. They are the negative value of portfolio returns (monthly) multiplied by 12. As the results show, both DeNN and DSNN are significantly outperforming the Equal weights and Random weights baselines in VaRs minimization. Besides, compared to their corresponding benchmarks, the DeNN and the DSNN also achieve a great improvement. For instance, comparing to LRM, the DeNN reduces the VaR at 95% when optimizing $CVaR_{95\%}$ by 17.45%, which corresponds to a decrease of annualized loss from 36.27% to 29.94%. As for DSNN, the results validate the effectiveness of our method. It shows that it is effective to connect SNN and DNN to help the training and prediction of SNN. Specifically, comparing to simple SNN, the DSNN reduces the VaR at 95% when optimizing VaR$_{95\%}$ by 6.26%, which corresponds to a decrease of annualized loss from 44.40% to 41.62%. Last but not the least, the DeNN model, which has a low degree of freedom and is quite simple in design, outperforms the DSNN model in most scenarios, which has a relatively high degree of freedom and is sophisticated in design.

| Confidence | 99% | **95%** | 90% | 85% | 80% |
|---|---|---|---|---|---|
| **DeNN** | **0.5810** | **0.3753** | **0.2731** | **0.2031** | **0.1648** |
| LRM | 0.6613 | 0.4364 | 0.3494 | 0.2811 | 0.2315 |
| **DSNN** | 0.6462 | **0.4162** | **0.3391** | **0.2730** | **0.2290** |
| SNN | **0.6449** | 0.4440 | 0.3537 | 0.2863 | 0.2337 |
| Equal | 0.6505 | 0.4740 | 0.3807 | 0.3139 | 0.2582 |
| Random | 0.7865 | 0.5213 | 0.3911 | 0.3121 | 0.2592 |

Table 2: VaRs at different confidence (optimizing $VaR_{95\%}$)

| Confidence | **99%** | 95% | 90% | 85% | 80% |
|---|---|---|---|---|---|
| **DeNN** | **0.6185** | **0.3590** | **0.2679** | **0.2087** | **0.1684** |
| LRM | 0.6419 | 0.4245 | 0.3276 | 0.2685 | 0.2239 |
| **DSNN** | 0.6655 | **0.4415** | **0.3458** | **0.2796** | **0.2280** |
| SNN | **0.6570** | 0.4444 | 0.3459 | 0.2854 | 0.2369 |
| Equal | 0.6505 | 0.4740 | 0.3807 | 0.3139 | 0.2582 |
| Random | 0.7865 | 0.5213 | 0.3911 | 0.3121 | 0.2592 |

Table 3: VaRs at different confidence (optimizing $VaR_{99\%}$)

| Confidence | 99% | **95%** | 90% | 85% | 80% |
|---|---|---|---|---|---|
| **DeNN** | 0.6649 | **0.2994** | **0.1992** | **0.1430** | **0.0972** |
| LRM | **0.5601** | 0.3627 | 0.2580 | 0.1976 | 0.1588 |
| **DSNN** | 0.5484 | **0.3603** | **0.2635** | **0.2159** | **0.1770** |
| SNN | 0.5993 | 0.3753 | 0.2821 | 0.2296 | 0.1951 |
| Equal | 0.6505 | 0.4740 | 0.3807 | 0.3139 | 0.2582 |
| Random | 0.7865 | 0.5213 | 0.3911 | 0.3121 | 0.2592 |

Table 4: VaRs at different confidence (optimizing $CVaR_{95\%}$)

| Confidence | **99%** | 95% | 90% | 85% | 80% |
|---|---|---|---|---|---|
| **DeNN** | 0.6034 | **0.3247** | **0.2223** | **0.1685** | **0.1253** |
| LRM | **0.5805** | 0.3624 | 0.2721 | 0.2099 | 0.1696 |
| **DSNN** | **0.6033** | 0.3768 | 0.2834 | 0.2326 | 0.1909 |
| SNN | 0.6367 | 0.3979 | 0.3027 | 0.2451 | 0.2061 |
| Equal | 0.6505 | 0.4740 | 0.3807 | 0.3139 | 0.2582 |
| Random | 0.7865 | 0.5213 | 0.3911 | 0.3121 | 0.2592 |

Table 5: VaRs at different confidence (optimizing $CVaR_{99\%}$)

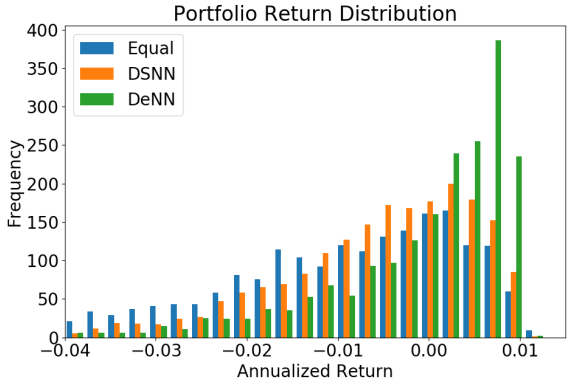

Figure 2: Portfolio Return Distribution (optimizing CVaR$_{95\%}$)

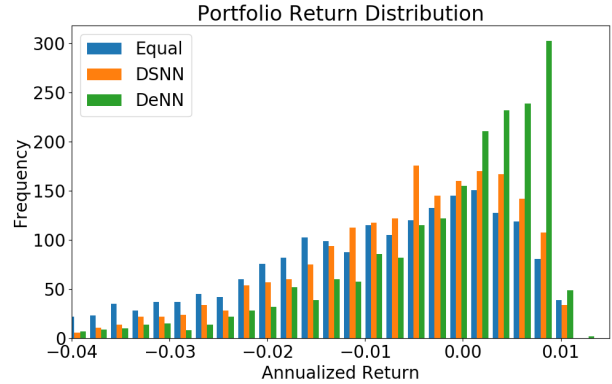

Figure 3: Portfolio Return Distribution (optimizing CVaR$_{99\%}$)

According to the tables, the VaRs at other confidence levels show considerable improvements as well. Therefore, we illustrate the histograms of 2,000 portfolios' realized returns for different weights generating methods. For simplicity and clarity, only DeNN, DSNN and Equal weights are plotted in Figure 2-3. Since the frequency of realized returns below -0.04 is quite low, we set the x-axes in the [-0.040, 0.015] interval to zoom in for comparisons. As shown in Figure 2-3, both the DeNN and the DSNN method skews the distribution of the realized portfolio returns to much higher levels. These results demonstrate that, exploiting the deep neural networks and optimization through Monte-Carlo simulations, our methods can significantly reduce the VaRs of a loan portfolio.

## 5 Literature Review

In literature, works about P2P lending mainly fall into two categories: default rate prediction and portfolio selection.

Predicting the default rates of loans is a classic application in data mining and a long-standing problem in finance. While logistics regression remains the most widely used method [Zhao et al., 2014; Guo et al., 2016], various machine learning techniques have been employed in previous works, such as Neural Networks [Byanjankar et al., 2015], Random Forests [Malekipirbazari and Aksakalli, 2015], GBDT [Zhao et al., 2016], XGBoost Decision Tree [Xia et al., 2017], etc. These models obtain good results on predicting the default rates of loans. However, a drawback of these methods is that they don't answer the question when will a loan default, which is significant in analyzing a loan's profitability. Therefore, survival analysis is employed in loans' defaults prediction [Banasik et al., 1999; Baesens et al., 2005]. Recently, Liao and Ahn [2016] design a deep neural network with survival analysis to predict the failure of hard drives. They achieve significant improvements comparing with traditional survival analysis model.

Another topic related is portfolio selection [Markowitz, 1952; Zhao et al., 2014; Guo et al., 2016; Zhao et al., 2016]. Unlike their papers, we use VaR/CVaR as our optimization objectives. Besides, Bitvai and Trevor [2015] is another important work in P2P lending. Since their concern is the loan rate determined by lenders' biddings, this work is not in line with ours. There are also some important works focusing on combining machine learning and optimization [Queipo et al., 2015; Lombardi et al., 2017]. They are related but our focuses are not the same.

## 6 Conclusions and Discussions

In this paper, we propose DeNN and DSNN with Monte-Carlo simulations to minimize the VaR and CVaR of loan portfolios. The experiments on real-life data demonstrate that our models can significantly reduce the VaR at different confidence levels.

There are several essential features that differentiate our paper from others. First of all, minimizing the VaR/CVaR of a P2P loan portfolio is a less touched problem with practical significance. And compared to several benchmarks, our methods achieve the-state-of-the-art performance on this problem. Second, we propose to use an 'expert' neural network that is a neural network trained exclusively for a certain task to help the training of another neural network. Third, to the best of our knowledge, this work is the first to use deep neural networks to predict the return distributions of P2P loans. These innovative applications of neural networks in finance make our task unique. Last but not the least, the DeNN model, which has a low degree of freedom and is quite simple in design, outperforms the DSNN model in most scenarios, which has a relatively high degree of freedom and is sophisticated in design. This result is interesting and could be insightful in guiding the use of neural networks in practice.